\documentclass[10pt,a4paper,final]{article}
\usepackage[utf8]{inputenc}
\usepackage{amsmath}
\usepackage{amsfonts}
\usepackage{amssymb}
\usepackage{graphicx}
\usepackage{url}

\begin{document}
\title{Model-driven Engineering of Safety and Security Systems: A Systematic Mapping Study}
\author{{\bf Atif Mashkoor}\\ Software Competence Center Hagenberg GmbH, Austria\\firstname.lastname@scch.at\\
{\bf Alexander Egyed}\\ Johannes Kepler University Linz, Austria\\firstname.lastname@jku.at\\
{\bf Robert Wille} \\ Johannes Kepler University Linz, Austria\\
firstname.lastname@jku.at
}

\maketitle
\begin{abstract}
This paper presents a systematic mapping study on the model-driven engineering of safety and security concerns in systems. Integrated modeling and development of both safety and security concerns is an emerging field of research. Our mapping study provides an overview of the current state-of-the-art in this field. Through a rigorous and systematic process, this study carefully selected 95 publications out of 17,927 relevant papers published between 1992 and 2018. This paper then proposes and answers several relevant research questions about frequently used methods, development stages where these concerns are typically investigated in, or application domains. Additionally, we identify the community’s preference for publication venues and trends. 
\end{abstract}

\section{Introduction}
\label{sec:intro}
The notion of (functional) safety is defined as a freedom from risk of damage to user, property or environment, and correct operation of a system in response to its inputs. Security, on the other hand, is the prevention of illegal access causing change and destruction of equipment, information and services~\cite{biro18a}. Traditionally, depending upon the domain and the background, software engineers mostly focused on designing systems, which are either safe or secure in nature. Typically, we find methods that investigate these concerns separately. While some exchange of ideas existed across these concerns, modern systems require their thorough, combined treatment since security concerns may affect safety and vice versa. This is particularly true for Cyber-Physical Systems or Internet of Things, whose components may be interconnected via the Internet and, hence, there is a good chance for safety concerns to be exposed and thus vulnerable to other types of failures caused by security attacks. Connected (and consequently exposed) safety systems must, therefore, be equipped with security mechanisms for protection against such attacks. Only recently, there has been a surge in research focusing on integrated modeling and development of safety and security systems~\cite{biro18a, wolf18a}.


Model-driven engineering~\cite{schmidt06a} is a development paradigm that focuses on creating models, which could be systematically transformed into (correct) pieces of software. The advantage is that developers can exclusively focus on modeling the problem rather than worrying about unnecessary and distracting implementation details. In that regards, model-driven engineering is appealing for addressing safety and security concerns where models play the integral role in describing and analyzing them.

Systematic mapping studies are meant to provide an overview of a research area through classification and counting contributions in relation to the categories of that classification. They involve searching the available literature in order to know what topics have been covered and where the corresponding papers have been published~\cite{petersen08a}. According to Kitchenham et al.~\cite{kitchenham10a}, the research questions in mapping studies are general as they aim to discover research trends (e.g., publication trends over time and topics covered in the literature). This is in contrast to systematic reviews, which intend to aggregate evidence and hence formulate a very specific goal (e.g., whether research results are practical and deployable for industry)~\cite{petersen15a}. The outcome of a mapping study is an inventory of publications on the selected topic mapped to a classification~\cite{wieringa05a}. 

In this paper, we present a systematic mapping study on model-driven engineering of safety and security systems. The overall aim of this study is to collect relevant state-of-the-art in this field of research. Besides that, we answer some important questions like what are frequently used methods and tools in this field, what are their applicable development stages, and in which application domains they have been evaluated. We also identify where the community prefers to publish research results and what are recent publication trends in this field. We prudently select 95 publications out of 17,927 relevant search results through a rigorous and systematic process. These publications were proven very helpful in answering a set of six judiciously crafted research questions, providing gainful insights, and identifying the directions for future research.    

The paper is organized as following: Sec.~\ref{sec:method} details the systematic mapping process including the research questions we investigate in this study. Sec.~\ref{sec:results} presents results of the mapping study. Sec.~\ref{sec:discussion} presents our analysis on the current state-of-the-art. Sec.~\ref{sec:threats} discusses the threats to validity of this study and the adopted mitigation strategies. The paper is concluded in Sec.~\ref{sec:conclusion}.

\section{Mapping study process}
\label{sec:method}
\paragraph{Time period}
We scope the time period of related studies published from 1992 to 2018. The earliest paper we could find in our mapping study was published in 1992, hence the starting time. We started this mapping study in December 2018, hence the ending time. Please note that we re-conducted the search in January 2019 in order to make sure that all results published until 2018 are included in our study.

\paragraph{Digital libraries}
Four digital libraries were used in this mapping study: ACM\footnote{\url{https://dl.acm.org}}, IEEE\footnote{\url{https://ieeexplore.ieee.org/Xplore/home.jsp}}, Springer\footnote{\url{https://link.springer.com}}, and Web of Science\footnote{\url{https://apps.webofknowledge.com}}. According to~\cite{chen10a}, these digital libraries are among the most popular sources in computer science and engineering that ensure a high coverage of potentially-relevant studies. Although Scopus\footnote{\url{https://www.scopus.com}} is also considered as an important source, it was not included in the study as it is not accessible in our institution. However, many venues indexed by Scopus are indexed by other included digital libraries as well. We did not include Google Scholar\footnote{\url{https://scholar.google.com}} in our mapping study as the search results of Google Scholar tend to be repetitive with respect to results from the included digital libraries, and its unique contribution to the search process is unclear~\cite{chen10a}.

\paragraph{Tool}
Conducting a systematic mapping study is a tedious and time consuming task. It usually involves search, collection, filtration, and classification of a huge amount of papers. Without a helping tool, this is a very difficult endeavor. In this work, we used Zotero~\cite{ahmed11a} and spreadsheets. These tools helped us in importing, organizing, and analyzing search results. 

\subsection{Research questions}
The goal of this mapping study (following the guidelines presented in~\cite{kitchenham07b, petersen08a, petersen15a}) is to discover what is the current state-of-the-art in the field of model-driven engineering of safety and security systems (and how it can be advanced in the future). This goal leads to following precise research questions (RQs):
\begin{itemize}
	\item RQ1: At which development stage the research was conducted?\\
	Rationale: Model-driven engineering is a multi-stage development process. We want to know at which development stage this research was conducted. This information can help us identify which development stages are susceptible to be focused more for the engineering of such systems.
	\item RQ2: Which methods and tools were employed during the research?\\
	Rationale: The use of methods and tools is inevitable during any research and development activity. This information can help us identify what are frequently used methods and tools for the engineering of such systems.
	\item RQ3: What is the classification of the contribution of the research?\\
	Rationale: Through this question, we want to investigate what is the contribution type of articles. According to Wieringa et al.~\cite{wieringa05a}, contribution types refer to determining the type of intervention being studied. This could be a process, method,
	model, tool, framework, etc.
	

	\item RQ4: In which domain(s) research results were evaluated?\\
	Rationale: Safety and security systems may belong to various application domains, e.g., railways, nuclear plants, and marine systems. By answering this question, we want to know what are the application domains in which the research results were evaluated. This information can help us identify which application domains have gained more interest of developers of such systems.
	\item RQ5: Where the research was published? \\
	Rationale: By answering this question, we want to find out whether researchers preferred to publish in journals, magazines, conferences, or workshops? Usually journals include more mature and concrete results, whereas conferences and workshops are targeted for timely discussion and early feedback. By answering this question, we can find out the maturity of results of this field. 
	\item RQ6: What is the research publication time-line and trend? \\
	Rationale: Time-lines and publication trends tell us about novelty and frequency of research. By answering this question, we can find out how the community is building around this topic. Is the topic a relatively new one, gaining popularity in recent years, or just phasing out? This information can help us determine the potential of this research topic.
\end{itemize}

\subsection{Papers search and screening}

\begin{figure}
	\centering
	\includegraphics[width=0.5\linewidth]{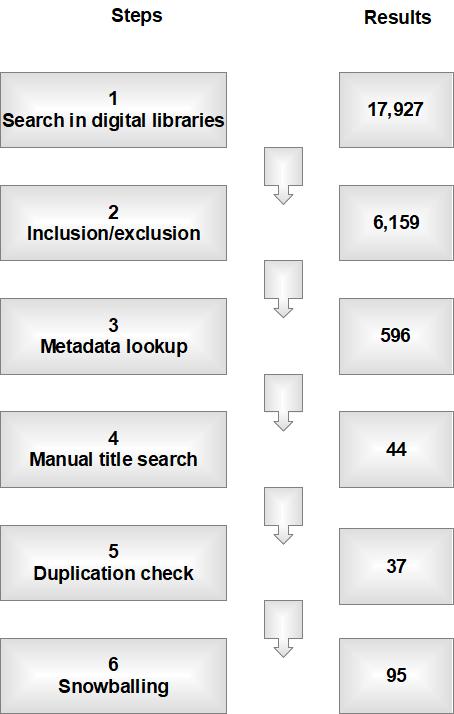}
	\caption{Steps for the search and selection process}
	\label{fig:approach}
\end{figure}

The mapping study was conducted in 6 steps as illustrated in Fig.~\ref{fig:approach}. 

\paragraph{Step 1 - Search in digital libraries}

The following query performed in aforementioned digital libraries produced 17,927 search results. Digital library-wise acquired results are shown in Tab.~\ref{tab:DLResults}.\\

\noindent \textit{(``model-driven'' OR ``model-based'') AND (``engineering'' OR ``development'') AND (``safety'' OR ``safe'') AND (``security'' OR ``secure'')}\\

The search terms were identified according to the study topic. 
Similar to Kitchenam et al.~\cite{kitchenham07b}, we adopted the PICO (Population, Intervention, Comparison, Outcomes) criteria to formulate the search terms. 

\begin{itemize}
	\item Population: According to Kitchenam et al.~\cite{kitchenham07b}, population may refer to a specific software engineering role, a category of software engineers, an application area, or an industry group. In our case, population is the terms about ``safety/safe'' and ``security/secure''.
	\item Intervention: According to Kitchenam et al.~\cite{kitchenham07b}, intervention may refer to a software methodology, a tool, a technology, or a procedure. In the context of this study, intervention includes terms ``model-driven'' or ``model-based''.
	\item Comparison: The comparison part is not applicable in this mapping study because this mapping study does not involve the comparison of model-driven and other types of approaches.
	\item Outcomes: Outcomes include the terms relevant to ``engineering'' or ``development'' activities.
\end{itemize}

We used the Boolean operator OR to join alternate words and synonyms in each part (i.e., population, intervention, outcomes), and the Boolean operator AND to join the terms from the three parts, respectively. 

\begin {table}
\caption {Digital library-wise results distribution} 
\label{tab:DLResults} 
\begin{tabular}{|c|c|c|}
	\hline 
	{\bf Digital library}& {\bf URL}  & {\bf \# of results}  \\ 
	\hline 
	ACM & \url{dl.acm.org}  &  193\\ 
	\hline 
	IEEE Xplore & \url{ieeexplore.ieee.org} & 238 \\ 
	\hline 
	Springer &  \url{link.springer.com} & 17,383 \\ 
	\hline 
	Web of Science & \url{apps.webofknowledge.com} &  113 \\ 
	\hline
	\bf{Total} & & {\bf 17,927}
	\\ 
	\hline
\end{tabular}

\end{table}

\paragraph{Step 2 - Inclusion and exclusion of results}

In order to make the study selection results objective, we defined selection criteria that were employed in the study selection process. The criteria is as following:

{\bf Inclusion criteria:}
\begin{itemize}
	\item Peer-reviewed studies published in conferences, workshops, journals, magazines or books,
	\item Studies classified as computer science publications, and
	\item Studies published in the English language.
\end{itemize}

{\bf Exclusion criteria:}
\begin{itemize}
	\item Studies published as courses or reference work entries,
	\item Studies published in disciplines other than computer science, 
	\item Studies published in languages other than English, and
	\item Studies presenting non-peer-reviewed results or gray literature.
\end{itemize}

In ACM and Web of Science digital libraries, the search process was simple. The basic search fields were enough to run the query, which required no further processing at this point. In the IEEE digital library, we had to use the advance search option. We wrote the query in the command search and obtained the results. The IEEE digital library also included courses on this topic in search results, which we excluded. We only included conferences, journals, magazines and early access articles. As shown in Tab.~\ref{tab:DLResults}, the Springer digital library produced the maximum amount of results. In order to focus on relevant results, we included only computer science publications, publications that were written in the English language, and publications that appeared as book chapters, conference papers or journal articles. We also excluded reference work entries from research results. This brought down the overall result count from 17,927 to 6,159 papers.

\paragraph{Step 3 - Meta-data lookup}

In this step, we carefully checked the available meta-data, e.g., keywords and abstracts, of results. We filtered out all those results, which were not explicitly focusing on safety and security. This brought down our results tally to 596.

\paragraph{Step 4 - Manual title search}

Despite the meta-data lookup, some publications could still be included in results, which are only remotely dealing with safety and security concerns. In order to ensure that the final result only includes high-quality relevant papers, we manually checked the title of each paper. We ensured that each paper title has something to do with both safety and security. As a result of this step, our result count became 44.

\paragraph{Step 5 - Check for duplicate results}

Once we individually checked results produced by each digital library, we merged all of them into a single repository. Since many digital libraries index same venues, the search results may be redundant. In order to have a list of only unique results, in this step we check the list of merged results for duplicates. All duplicates were consequently removed from the list of results and the publication count became 37.

\paragraph{Step 6 - Snowballing}

In this last step, we performed snowballing readings. In a snowballing reading, the lists of references of found papers are used to identify other relevant studies~\cite{wohlin14a}. We performed the backward snowballing, i.e., we went through lists of references of previously obtained papers and identified the potential relevant studies. For each paper identified for the possible inclusion, we applied the same criteria that were employed for the selection of papers in the first place. Then, we identified 58 further relevant studies in this step. After this, the final set of results reached the tally of 95 publications.

\subsection{Studies classification scheme}

The classification scheme used for this study follows a systematic process suggested by Petersen et al.~\cite{petersen08a, petersen15a}. We are using keywords as bases for studies classification. Initially, we read abstracts in order to find representative keywords and concepts. The set of extracted keywords from different studies are then unified in order to overview the nature and contribution of the research (e.g., as shown in Fig.~\ref{fig:methods}, different similar methods and tools are merged into a single category). This results into a set of categories which is representative of the underlying population. Sometimes meaningful keywords could not be extracted from the abstract alone. In such cases, either introduction and conclusion sections were studied, or complete papers were skimmed through. Upon selection of the final set of keywords, they are clustered and consequently used to form the categories. Where applicable, classifications were also based on the Software Engineering Body of Knowledge (SWEBOK)\footnote{\url{https://www.computer.org/education/bodies-of-knowledge/software-engineering}} structure (e.g., as shown in Fig.~\ref{fig:stages}, which are the main life-cycle activities of software engineering) or inspired from previous categorizations (e.g., as shown in Fig.~\ref{fig:category}, which is based on the work of Wieringa et al.~\cite{wieringa05a}). 

In order to reduce any bias, we followed an iterative strategy. Initially, the first author classified all studies. The classifications were then reviewed by the second author and corrected, where necessary. In case of a disagreement, the third author independently reviewed the classification and gave his judgment. The opinion of majority prevailed. Although all authors of this paper are senior and experienced researchers, given the fact that this step involves human judgment, the threat of bias cannot be completely eradicated.

\subsection{Data extraction and synthesis}
To answer the RQs, we extracted specific data from selected publications. Tab.~\ref{tab:extractedItems} describes data items that have been extracted in this mapping study. 

\begin {table}
\caption {Extracted research items} 
\label{tab:extractedItems} 
\begin{tabular}{|p{2cm}|p{4cm}|p{1.5cm}|}
	\hline 
	{\bf Item name}& {\bf Description}  & {\bf Relevant RQ}  \\ 
	\hline
	Development stage & At which development stage was the approach applicable? &  RQ1\\ 
	\hline
	Method/tool & What is the deployed method or tool? & RQ2\\  
	\hline 
	Contribution classification & What is the classification of contribution of the publication? & RQ3 \\ 
	\hline
	Application domain & Which application domain the study was applied to? & RQ4 \\ 
	\hline
	Publication type & In which publication type was the study published?  & RQ5a\\ 
	\hline
	Publication venue & In which venue was the study published? & RQ5b\\ 
	\hline
	Publication year & In which year was the study published? & RQ6\\ 
	\hline
\end{tabular}
\end{table}

Data synthesis targets to synthesize the extracted data to answer the RQs. The results of this task are discussed (also visually) in the following section.

\section{Mapping study results}
\label{sec:results}
In this section, we synthesize the extracted data to answer previously listed RQs. 

\subsection{Development stages (RQ1)}
\label{subsec:stages}

\begin{figure}
	\centering
	\includegraphics[width=0.9\linewidth]{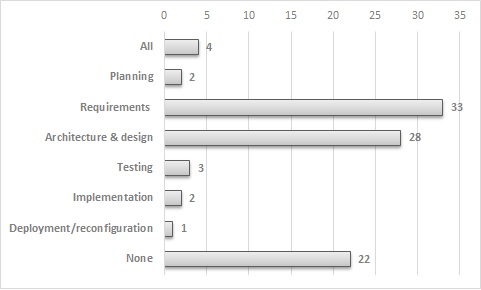}
	\caption{RQ1: Studies classification based on development stages}
	\label{fig:stages}
\end{figure}

As shown in Fig.~\ref{fig:stages}, only a few studies (4 out of 95) were covering the whole model-driven engineering spectrum, i.e., all development stages. Hassan et al.~\cite{hassan10a} were discussing the idea that how the Formal Analysis and Design for Engineering Security (FADES) approach can be used to support the model-based software engineering paradigm. Bloomfield et al.~\cite{bloomfield13a} use the structured safety cases approach to discuss the impact that security might have on an existing safety case. Apvrille et al.~\cite{apvrille16a} presented a similar idea based on the SysML-Sec environment covering all the development stages. However, all these studies did not use any application domain to evaluate the proposed approaches. The approach presented in~\cite{benyo16a}, on the other hand, discussed the design and development of a smart card application management infrastructure by specifying business and technological processes and associated security requirements.

Although the planning phase is very crucial for successful development of a system, very few studies (2 out of 95) have focused on this stage. In the first study, Park et al.~\cite{park16a} discuss how multi-agent systems and swarm intelligence can be exploited to boost counter-terrorism and public safety activities. In the second study, Park et al.~\cite{park16a} emphasize the idea of using cybersecurity considerations to make nuclear facilities even safer.   

As anticipated, most of the studies (33 out of 95) were focusing at the level of requirements. 14~\cite{elliot95a, eames99a, novak07a, poslad09a, sun09a, pietre10b, amthor11a, oates13a, apvrille15a, hessami15a, troubitsyna16a, troubitsyna16b, brunner17a, ponsard18a} out of those 33 were focusing exclusively on requirements modeling of such systems. 16 studies~\cite{fischer10a, monakova12a, monakova12b, raspotnig12a, dong12a, kornecki13a, bieber14a, brunel15a, li15a, taguchi15a, ponsard16a, pereira17a, vistbakka17a, pawlik18a, troubitsyna18a, sangchoolie18a} were focusing on both requirements modeling and analysis. A few studies were either focusing solely on requirements analysis~\cite{pietre10a, roudier15a} or requirements traceability~\cite{katta13a}.


The architecture and design stage is of paramount importance in the development of any system; safety and security systems are no exceptions. 28 out of 95 studies were focusing on this stage. 4 out of those studies 
~\cite{varrette09a, glasser10a, brunel14a, tverdyshev16a} were discussing architecture modeling. 10 studies~\cite{winther01a, preschern13a, steiner13a, kriaa14a, schmittner14a, woskowski14a, chen14a, macher15a, schmittner15a, subramanian16a} were discussing architecture analysis. 14 studies~\cite{glasser10b, jackson11a, banerjee12a, young13a, young14a, brunel14b, kriaa15b,  chen15a, cimatti15a, schmittner16a, martin17a, hazell17a, amorim17a, friedberg17a} were discussing how to make architectural design of systems both safe and secure through modeling and analysis.

Testing also plays a pivotal role in systems development life-cycle. We found 3 studies focusing on testing in our mapping study. While Sojka et al.~\cite{sojka14a} were explicitly focusing on the testing of safety and security requirements within the automotive domain, Shahir et al.~\cite{shahir11a, shahir12a} were focusing on test case generation for safety and security of marine systems.  

Development stages, such as implementation~\cite{kleidermacher12a, bagnara18a}, and deployment and reconfiguration~\cite{tariq18a}, were also mentioned in the literature, however, they were not the center of attention of researchers of this field.  

A large amount of studies (22 out of 95) were not focusing on any development stage at all. They were either comparing safety and security concepts, e.g.,~\cite{burns92a}, discussing how one can help achieving the other, e.g.,~\cite{brewer93a}, making similarities and dissimilarities explicit between the two, e.g.,~\cite{blanquart12a}, analyzing how the two concepts can cross-fertilize each other, e.g.,~\cite{pietre13a}, etc. 

\subsection{Methods and tools (RQ2)}
\label{subsec:methods}

\begin{figure}
	\centering
	\includegraphics[width=0.9\linewidth]{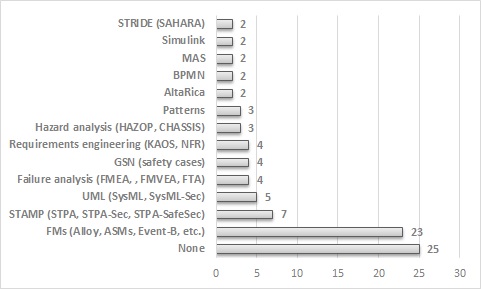}
	\caption{RQ2: Studies classification based on applied methods and tools}
	\label{fig:methods}
\end{figure}

Fig.~\ref{fig:methods} graphically depicts the frequency of the applied methods and tools. Some approaches consisted of more than one method/tool. The methods whose use has been confined to only one study are being skipped here for the sake of brevity. 

Our research shows that a multitude of methods and tools are being used in this field but none clearly stands out. Although there is an observable tendency among the community to use formal methods for such kind of engineering activities (23 studies are conducted using formal methods), yet no particular formal method can be classified as the method of choice. Among more frequently used formal methods, the use of Abstract State Machines has been found in five studies~\cite{glasser10a, glasser10b, jackson11a, shahir11a, shahir12a}, whereas the use of Event-B~\cite{troubitsyna16a, vistbakka17a, troubitsyna18a} and Alloy~\cite{brunel15a, brunel14a, brunel14b} has been mentioned in three studies apiece. However, please note that all applications of Abstract State Machines and Event-B method stemmed from single groups and targeted towards marine and industrial control systems, respectively. 

The second most widely used set of techniques, after formal methods, was based on STAMP and its variants, i.e., STPA, STPA-Sec, and STPA-SafeSec (7 studies). Systems-Theoretic Accident Model and Processes (STAMP)~\cite{leveson03a} is an accident causality model based on systems theory and systems thinking. Systems-Theoretic Process Analysis (STPA)~\cite{ishimatsu10a} is a powerful hazard analysis technique based on STAMP. STPA-Sec~\cite{young14a} is a system-theoretic process analysis method explicitly focusing on security issues.  STPA-SafeSec~\cite{friedberg17a} is an analysis methodology for both safety and security. The use of STAMP is mentioned in~\cite{troubitsyna16a}. The use of STPA is mentioned in~\cite{pereira17a}. The use of STPA-Sec is mentioned in~\cite{young13a, young14a, schmittner16a}. The use of STPA-SafeSec is mentioned in~\cite{friedberg17a}. 

The use of UML and its variants, i.e., SysML and SysML-Sec is also relatively popular in this domain and has been found in 5 studies. The use of Unified Modeling Language (UML) is mentioned  in~\cite{raspotnig12a}. The use of Systems Modeling Language (SysML) has been mentioned in~\cite{oates13a}. The use of SysML-Sec (an extended version of the SysML language to design safe and secure embedded systems) has been found in~\cite{apvrille15a, roudier15a, apvrille16a} .    

Another approach relatively popular in this domain is based on Goal Structuring Notation (GSN) and safety cases. The use of these notations has been found in 4 studies. The use of GSN is mentioned in~\cite{preschern13a, troubitsyna16b, martin17a} and the use of safety cases is mentioned in~\cite{bloomfield13a, troubitsyna16b}.

Goal-oriented requirements engineering approaches, such as KAOS or NFR, also play an important role in this domain. Their use has been found in 4 studies. The use of Knowledge Acquisition in Automated Specification (KAOS) -- a goal-oriented requirements engineering approach -- has been found in~\cite{ponsard16a, ponsard18a}. The use of the Non-Functional Requirements (NFR) approach -- a goal-oriented technique that can be applied to determine the extent to which specific objectives are achieved by a design -- has been found in~\cite{kornecki13a, subramanian16a}.    

Failure analysis is the process of collecting and analyzing data to determine the cause of a possible failure in a system. Failure analysis methods (e.g., FMEA, FMVEA, and FTA) are also commonly used in the engineering of safe and secure systems. Their use has been found in 4 studies. The use of Failure Mode and Effects Analysis (FMEA) is mentioned in~\cite{winther01a}. The use of Failure Mode, Vulnerabilities and Effects Analysis (FMVEA) -- a variant of FMEA -- has been found in~\cite{schmittner14a, schmittner15a}. The use of Fault Tree Analysis (FTA) has been found in~\cite{steiner13a}. An approach very similar to failure analysis is hazard analysis (e.g., HAZOP and CHASSIS). The use of hazard analysis approaches has been found in 3 studies, however, 2 of them were applied in combination with traditional failure analysis methods. The use of Hazard and Operability Study (HAZOP) in combination with FMEA is found in~\cite{winther01a} and in combination with Combined Harm Assessment of Safety and Security for Information Systems (CHASSIS) has been found in~\cite{katta13a}. The use of CHASSIS in combination with FMVEA has been found in~\cite{schmittner15a}. 

Some researchers have also proposed different patterns in this domain, i.e., architectural safety patterns including security considerations~\cite{preschern13a}, safe \& sec case patterns~\cite{taguchi15a}, and systems engineering patterns for interlinking safety and security~\cite{amorim17a}. 

There are many other methods and tools which have been used less frequently as compared to the aforementioned methods and tools. Following are the methods whose use has been found in 2 studies apiece.  The use of AltaRica -- a high-level language designed for the modeling of systems -- has been mentioned in~\cite{bieber14a, brunel14b}. The use of Business Process Model and Notation (BPMN) -- a graphical representation for specifying business processes -- has been found in~\cite{monakova12a, monakova12b}. The use of Multi-Agent Systems (MAS) has been found in~\cite{poslad09a, park12a}. The use of Simulink -- a graphical programming environment for modeling, simulating and analyzing multi-domain dynamical systems -- has been found in~\cite{tariq18a, sangchoolie18a}. The use of the Spoofing, Tampering, Repudiation, Information Disclosure, Denial of Service (DoS), Elevation of Privilege (EoP) (STRIDE) approach -- a structured way to find security threats -- has been found in~\cite{preschern13a, macher15a}. 

A large number of found studies (25) did not employ any particular method or tool in the conducted research. They were either characterizing the differences between safety and security, e.g.,~\cite{burns92a}, comparing the two approaches~\cite{raspotnig13a}, stressing the need of their integration, e.g.,~\cite{eames99a}, or demonstrating how they could possibly complement each other, e.g.,~\cite{brewer93a}. 

\subsection{Contribution classification (RQ3)}
\label{subsec:category}

\begin{figure}
	\centering
	\includegraphics[width=0.9\linewidth]{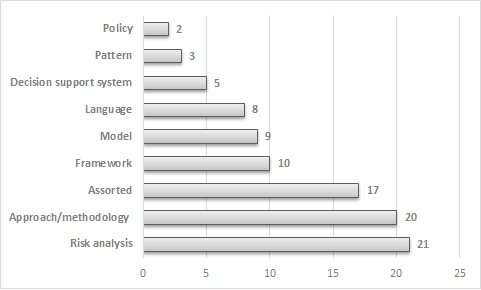}
	\caption{RQ3: Studies classification based on contributions}
	\label{fig:category}
\end{figure}

As shown in Fig.~\ref{fig:category}, the contribution of a study can be classified into multiple categories. Most of the researchers were interested in deploying model-driven approaches for risk-related activities of safe and secure systems. 21 out of 95 studies focused on this aspect. While majority of researchers explicitly focused on risk analysis~\cite{winther01a, aven07a, young13a, brunel14b, brunel14a, kriaa14a, schmittner14a, young14a, macher15a, schmittner15a, kriaa15b, schmittner16a, martin17a, friedberg17a}, some also focused on risk assessment~\cite{bloomfield13a, katta13a, chen14a}, risk communication~\cite{fruth14a}, risk management~\cite{woskowski14a, hazell17a}, and risk modeling~\cite{sangchoolie18a}.  

20 out of 95 publications~\cite{poslad09a, pietre10b, kornecki13a, steiner13a, bieber14a, brunel15a, sojka14a, nostro14a, chen15a, ponsard16a, troubitsyna16a, park16a, subramanian16a, troubitsyna16b, pereira17a, vistbakka17a, brunner17a, ponsard18a, troubitsyna18a, tariq18a} proposed an approach or a methodology based on an already existing method or tool. Their aim was to adapt an existing technology in such a way that it becomes exploitable for model-driven engineering of safety and security systems. 

17 out of 95 studies belonged to the assorted class. We put those studies in this class which could not be classified otherwise. Most of these studies were either comparative or conceptual, i.e., some of them were comparing safety and security concepts, e.g.,~\cite{burns92a}, some of them were eliciting similarities and dissimilarities between the two concepts, e.g.,~\cite{blanquart12a}, some of them were arguing why the two concepts are vital for Cyber-Physical Systems and Internet of things, e.g.,~\cite{wolf17a}, some of them were aligning the two concepts with each other, e.g.,~\cite{skoglund18a}, etc.

10 out of 95 studies presented a framework, which could be useful in various phases of model-driven engineering of safety and security systems. These frameworks were based on either formal methods~\cite{sun09a, banerjee12a, dong12a, li15a}, SysML~\cite{oates13a}, or other various methods and tools~\cite{varrette09a, pietre10a, park12a, schmittner15b, benyo16a}.  

9 out of 95 studies presented a model. The model could either be related to the collaborative modeling of safety and security properties~\cite{robinson-mallett14a}, a life-cycle model~\cite{novak07a, kleidermacher12a, kanamaru17a}, a process model~\cite{eames99a, raspotnig12a}, a reference model~\cite{pawlik18a}, or a business process model~\cite{monakova12a, monakova12b} for the development of safety and security systems.

8 out of 95 studies presented a language for the model-driven engineering of safety and security systems. The proposed languages were either formal languages~\cite{hassan10a, fischer10a, cimatti15a}, graphical modeling languages~\cite{apvrille15a, roudier15a, apvrille16a}, programming languages~\cite{bagnara18a}, or a language for presenting mishaps~\cite{stoneburner06a}.

Some researchers also presented some decision support systems~\cite{glasser10a, glasser10b, jackson11a, shahir11a, shahir12a}, patterns~\cite{preschern13a, taguchi15a, amorim17a}, and policies~\cite{amthor11a, tverdyshev16a} for the model-driven engineering of safety and security systems. 

\subsection{Evaluation domains (RQ4)}

\begin{figure}
	\centering
	\includegraphics[width=0.9\linewidth]{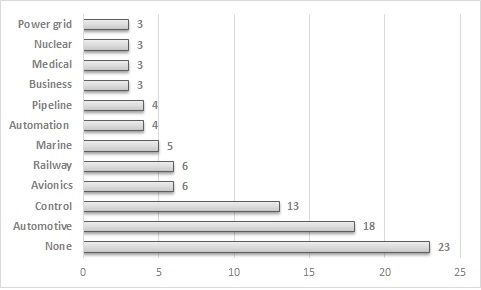}
	\caption{RQ4: Studies classification based on evaluation domains}
	\label{fig:domains}
\end{figure}

Fig.~\ref{fig:domains} graphically depicts the frequency of evaluation domains. Some publications used more than one domain for evaluation purposes. Domains confined to only one study are being skipped here for the sake of brevity.

Researchers dealing with safety and security were mostly interested in evaluating their proposed methodologies and tools in the automotive domain. 18~\cite{steiner13a, schmittner14a, sojka14a, robinson-mallett14a, apvrille15a, macher15a, roudier15a, schmittner15b, schmittner15a, ponsard16a, schmittner16a, schoitsch16a, amorim17a, martin17a, brunner17a, huber18a, skoglund18a, sangchoolie18a} studies were applied to automotive systems. 

After automotive systems, researchers of this field were mostly interested in applying their methods and tools in control systems. 13 studies were applied to control systems. Control systems included general-purpose control systems~\cite{young13a, kleidermacher12a}, air traffic control systems~\cite{eames99a, raspotnig12a, katta13a}, access control systems~\cite{amthor11a}, industrial control systems~\cite{oates13a, kriaa15a, kriaa15b, troubitsyna16a, tverdyshev16a, vistbakka17a}, and network control systems~\cite{troubitsyna18a}.  

After control systems, avionics and railway domains were most attractive for researchers of this topic. Both domains were used as a testbed in 6 studies apiece. The avionics domain has been featured in~\cite{banerjee12a, bieber14a, brunel14b, brunel14a, nostro14a, pereira17a}. Whereas, the railway domain has been featured in~\cite{winther01a, dong12a, hessami15a, pawlik17a, pawlik18a, ponsard18a}.   

The use of marine systems as an evaluation domain has been mentioned in 5 studies~\cite{glasser10a, glasser10b, jackson11a, shahir11a, shahir12a}. However, an interesting point to note is that all of these publications stemmed from a single group applying a particular method: Abstract State Machines.

Automation and pipeline systems were mentioned in 4 studies each. Automation systems were further classified into building automation systems~\cite{novak07a, sun09a}, electrical substation automation systems~\cite{preschern13a}, and industrial automation systems~\cite{fruth14a}. Pipeline systems, on the other hand, were mainly dealing with the oil industry~\cite{kornecki13a, kriaa14a, li15a, subramanian16a}.

Business, medical, nuclear, and power grid systems have been used as an evaluation domain in 3 publications each. The use of business systems, mainly enterprise resource planning systems, has been mentioned in~\cite{fischer10a, monakova12a, monakova12b}. The use of medical systems has been mentioned in~\cite{varrette09a, banerjee12a, woskowski14a}. The use of nuclear systems has been mentioned in~\cite{pietre10a, chen14a, park16a}. The use of power grid systems has been mentioned in~\cite{pietre10a, friedberg17a, tariq18a}.

The use of defense~\cite{cimatti15a}, fire detection~\cite{brunel15a}, rescue~\cite{park12a}, road transportation~\cite{chen15a}, smart card~\cite{benyo16a}, and virtual organization~\cite{poslad09a} systems has been mentioned only once in the found literature. 

As aforementioned in Sec.~\ref{subsec:methods}, many found studies did not employ any particular method or tool in their research. These studies were either characterizing the differences between safety and security or stressing the need of their integration. Naturally, such comparative or road-map studies were not always subject to evaluation. Consequently, a large number of studies (23) we found were not evaluated on any particular domain. 

\subsection{Publication types and venues (RQ5)}

\begin{figure}
	\centering
	\includegraphics[width=0.9\linewidth]{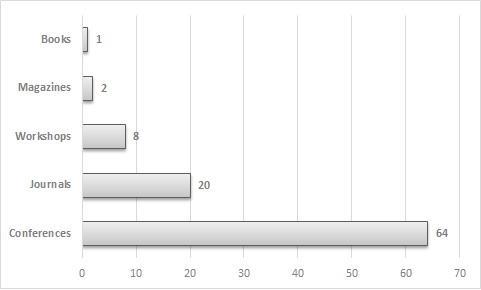}
	\caption{RQ5a: Studies classification based on publication types}
	\label{fig:venues}
\end{figure}

\paragraph{Publication types (RQ5a)}
In this study, only peer-reviewed publications (including books, journals, magazines, conferences and workshops) were considered. Fig.~\ref{fig:venues} provides an overview of the distribution of studies between these venues. An overwhelming majority of studies (64/95) were published in conferences followed by journals and workshops, respectively. 

\begin{figure}
	\centering
	\includegraphics[width=0.9\linewidth]{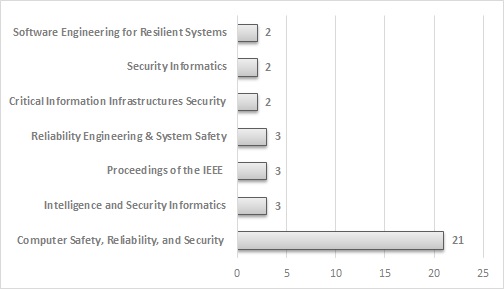}
	\caption{RQ5b: Studies classification based on top publication venues}
	\label{fig:topvenues}
\end{figure}

\paragraph{Publication venues (RQ5b)}
Looking at mapping study results, it was clear which venues were mostly targeted by researchers of this domain. The top seven venues for this domain are shown in Fig~\ref{fig:topvenues}. The most favorite venue of researchers of this topic is undoubtedly the Conference on Computer Safety, Reliability, and Security (Safecomp). 21~\cite{eames99a, winther01a, bieber14a, brunel14a, fruth14a, kriaa14a, schmittner14a, woskowski14a, macher15a, cimatti15a, schmittner15b, taguchi15a, ponsard16a, schmittner16a, troubitsyna16a, amorim17a, martin17a, pereira17a, huber18a, skoglund18a, troubitsyna18a} out of 95 studies are published on this venue. Conference on Intelligence and Security Informatics (ISI)~\cite{glasser10b, jackson11a, shahir11a}, Proceedings of the IEEE~\cite{banerjee12a, wolf17a, tariq18a}, and Journal on Reliability Engineering \& Systems Safety~\cite{aven07a, pietre13a, kriaa15a} contained 3 publications each. Conference on Critical Information Infrastructures Security (CRITIS)~\cite{tverdyshev16a, chockalingam17a}, Journal on Security Informatics~\cite{park12a, shahir12a}, and Workshop on Software Engineering for Resilient Systems (SERENE)~\cite{bloomfield13a, vistbakka17a} hosted 2 papers apiece. All other venues had only 1 publication each.

\subsection{Publication time-line and trend (RQ6)}

\begin{figure}
	\centering
	\includegraphics[width=0.9\linewidth]{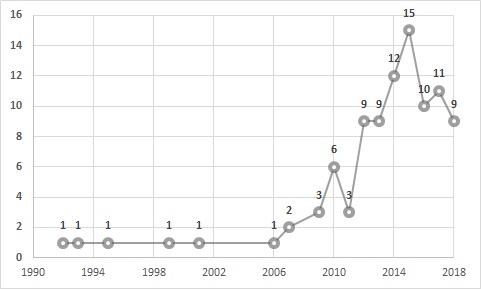}
	\caption{RQ6: Studies classification based on publication time-line and trend}
	\label{fig:trend}
\end{figure}

Fig.~\ref{fig:trend} shows the time-line and trend of publications in this area. As per our findings, the first study explicitly focusing on safety and security together was published in 1992~\cite{burns92a}. While the interest in this area was moderately increasing until 2011, a significant increase can be observed in 2012 onwards reaching its top in 2015. Since then, like a typical hype cycle, the community is perhaps slowly climbing the ``slope of enlightenment'' towards the ``plateau of productivity''. Nonetheless, the increase in the number of publications indicates that the area is  considered highly relevant by the software engineering research community.

\section{Discussion}
\label{sec:discussion}
Regarding RQ1, we have found that majority of researchers are working at the level of requirements or architecture. Modeling and analysis activities are the primary focus at both these levels. Only few researchers are considering the whole model-driven engineering spectrum (i.e., all development life-cycle activities). While modeling and analysis of requirements and designs are important activities, it is also imperative to ensure that these models are eventually translated into implementations as seamlessly as possible. Detailed works showing such transformations are currently missing from the state-of-the-art and worth exploring in the future. Another important point we have observed in found studies is that testing is not a primary focus of researchers of this field. Although the code generated through a rigorous development process is, in principle, already verified and validated, this is not enough in case of critical systems. For such systems, the generated code also needs to be tested~\cite{bonfanti18a}. In our opinion, researchers of this field should also give priority to testing as it uncovers different set of problems than those found in earlier stages of development, e.g., if the code is later manually modified in order to introduce further implementation details, the designer can use tests in order to check that no faults are introduced inadvertently.  

Regarding RQ2, we have found out that no single method or tool is prevalent in this domain. Although the use of formal approaches is common (which, by the way, makes perfect sense given the critical nature of safety and security systems), no particular formal method stands out. Formal methods, such as Abstract State Machines or Event-B, have been used in design and development of many systems. However, the use of these methods often stems from particular groups. The use of the STAMP method -- originally proposed for the safety domain -- also looks promising in this field and several variants of STAMP have recently been proposed to extend its capability towards security systems. However, it needs to be applied to more domains and projects before its actual suitability for safety and security systems can be truly evaluated. Additionally, the current use of this method is also confined to modeling and analysis of requirements and design artifacts. In the future, application of this method (by extension) to other stages of development could be an interesting topic of research.  The use of graphical modeling languages, such as UML or SysML, is certainly lacking behind in this field. Even the applied work is mostly concentrated on modeling and analysis of requirements. Given the potential of these languages, there could be a niche for their users to demonstrate their effectiveness through their widespread applications to safety and security systems.

Regarding RQ3, we have found out that most researchers were interested in risk analysis of such systems. Risk analysis is a crucial activity in the domain of safety. This becomes even more crucial when safety is integrated with security. Variety of methods were used for risk analysis and mostly researchers worked at the architecture and design level. While hazard analysis (e.g., HAZOP) and failure analysis (e.g., FMEA or FTA) are already established methods for risk analysis, new approaches, such as FMVEA, STPA-SafeSec or SAHARA, are also emerging recently. Working towards maturity and improvement of these approaches by their further application to new domains and projects is also an exciting research topic. Another interesting observation we made was that most researchers are extending the capabilities of existing methods and tools to solve the challenges of this field (e.g., FMVEA is based on well-established FMEA or STPA-SafeSec is based on popular STPA) rather than presenting new frameworks and languages. Of course, new pertinent frameworks (e.g., SAFESCALE~\cite{varrette09a}) or languages (e.g., FADES~\cite{hassan10a}) are also surfacing, but relatively low in number.    

Regarding RQ4, we have found out that most researchers used the automotive domain to evaluate their results. This is consistent with the emerging phenomenon of autonomous driving where both safety and security play equally critical roles. However, the prominence of research in model-driven engineering for the automotive domain predates autonomous driving and has more to do with the adoption of this paradigm by the automotive industry~\cite{broy12a}. Control systems were also a favorite testbed for the evaluation of such systems.   

Regarding RQ5, we have found out that most researchers preferred to publish their results in conferences and especially in SAFECOMP (International Conference on Computer Safety, Reliability and Security). Articles appearing in journals were not only less in numbers but also distributed among different venues. The numbers indicate that this research field is still (relatively) young and evolving.

Regarding RQ6, we have found out that the interest of community is increasing in this research field. In past few years, more and more publications are appearing explicitly focusing on the model-driven engineering of safety and security systems. 

We have observed in our research that a significant amount of publications did not mention any development stage, method, or evaluation domain in their results. This is mainly due to the fact that these publications were stressing the need of joint modeling and development of safety and security by either comparing the two concepts, discussing how one can help achieving the other, or analyzing how the two concepts can cross-fertilize each other. Naturally, such conceptual and road-map studies were not subject to classification in respective RQs.

In the end, we would like to express that, in our opinion, state-based formal methods\footnote{a primer on state-based formal methods is available in~\cite{mashkoor18a}} supporting the ``correct-by-construction'' development approach seem to be quite suitable for engineering of such kind of systems: they cover all stages of development life-cycle, variety of modeling and analysis tools are available at the disposal of developers, quality assurance is embedded within the development process, there is a support for translation of requirements and design artifacts into correct pieces of software, etc. The catch is that state-based formal models may be opaque, i.e., hard to read and write for many stakeholders~\cite{kossak16a, kossak14c}. In this case, such developments can be augmented by using graphical modeling notations such as UML or SysML. This not only provides cross-fertilization among various modeling tools but also enables developers to harness the true potential of each tool at the suitable development stage. Some tools (e.g., UML-B~\cite{snook06a}) and approaches (e.g., KAOS-Event-B~\cite{mashkoor10a}) already exist and worth exploring in this regard. As far as risk analysis is concerned, which generally does not fall within the jurisdiction of formal methods, further research is required towards its harmonization with formal methods such as shown by Khan et al.~\cite{khan18a}. Another problem with state-based formal methods is that while they offer effective tools for verification and validation, the support for automatic code generation is far from desirable~\cite{mashkoor16c}. These methods can, in principle, generate code artifacts from models, however, the generated code needs a lot of manual post processing. This may introduce some inconsistencies or errors in code, which may, in turn, compromise the integrity of previously applied rigorous quality assurance process. This also makes systems susceptible to extensive testing, which is already a weak link in the development chain of such systems. Hence, future methods for model-driven engineering of safety and security systems need to offer better tools and methodologies, especially for code generation and testing, respectively.

\section{Threats to validity}
\label{sec:threats}
There is always a threat of  validity for such kind of research. We also face a number of threats in our systematic mapping process, which we discuss as following.

The first threat is related to research questions: are they the right kind of questions we should be asking? To minimize this threat, we judiciously crafted the questions in alignment with the overall aim of this work after having several internal discussions. The final set of research questions indeed reflect the goals of our work.

The second threat is related to the terms used in search queries. To minimize this threat, we adopted the PICO (Population, Intervention, Comparison, Outcomes) criteria~\cite{kitchenham07b} to formulate the search terms. The selected terms unequivocally represents the goals of our work. An associated issue corresponds to the frequently used acronyms for model-based/driven engineering. Although the query used did not explicitly include related acronyms, such as MDE (model-driven engineering), MDD (model-driven development), or MBSE (model-based software engineering), this would not result in missing relevant articles because such information is usually available (or redundant) in meta-data, e.g., keywords or index terms, hence accessible.

The third threat is regarding the source of the data. We used four digital libraries as a primary source for this research. All selected digital libraries are well known in the computer science discipline for including most relevant results~\cite{dyba07a}. Although Scopus is also considered as an important source, it was not included in the study due to its inaccessibility at our research institute. However, many venues indexed by Scopus are already indexed by other considered digital libraries. Additionally, Wohlin et al.~\cite{wohlin13a} state that having a larger set of papers is not necessarily better for mapping studies. The important thing is that found studies are a good representation of the population, which we ensured in this study by adopting a rigorous paper selection process.

The fourth threat is regarding the quality assessment. As found by Petersen et al.~\cite{petersen15a}, quality assessment is not common in mapping studies. This is also consistent with suggestions of Kitchenham et al.~\cite{kitchenham10a}, which state that quality assessment is not essential for mapping studies as their overall aim is to give a broad overview of the topic area. However, despite these observations, we have adopted a rigorous process for inclusion/exclusion and classification of papers, which ensures that only high-quality related papers are selected as primary studies.

\section{Conclusion}
\label{sec:conclusion}
In this paper, we have presented a systematic mapping study on model-driven engineering of safety and security systems. Our mapping study provides an overview of the current state-of-the-art in this field. Through a rigorous and systematic process, this study carefully selected 95 publications out of 17,927 relevant search results, which proved very helpful in answering the judiciously crafted research questions like what are the frequently used methods and tools, what are the applicable development stages, and what are the evaluation domains. Additionally, we identified the community's preference for publication venues and publication trends. Based on analysis of selected studies, we indicated several avenues for future research.

The current state-of-the-art provides effective support for modeling and analysis of requirements and design of safety and security systems. However, the state-of-the-art needs to be advanced in order to offer better tools and methodologies especially for code generation and testing, respectively. Better integration of graphical modeling languages with conventional formal notations and better harmonization of rigorous methods and risk analysis approaches will also help in this regard. We also welcome more studies encapsulating the whole spectrum of model-driven engineering applied to safety and security systems.

In the future, 
we want to extend this study by asking qualitative questions like what is the maturity level of the presented contribution, how useful it is for the given task, which impetuses are required as input, and whether the contribution is applicable at the design time (static) or the runtime (dynamic). 

\section*{Acknowledgement}
The research presented in this technical report has been partly funded by the LIT Secure and Correct Systems Laboratory and the Austrian Ministry for Transport, Innovation and Technology (BMVIT), the Federal Ministry for Digital and Economic Affairs (BMDW), and the Province of Upper Austria in the frame of the COMET - Competence Centers for Excellent Technologies - program managed by the Austrian Research Promotion Agency FFG.

\end{document}